# Dynamic *d*-symmetry Bose condensate of a planar-large-bipolaron-liquid in cuprate superconductors

*David Emin, Department of Physics and Astronomy, University of New Mexico, Albuquerque, New Mexico 87131 USA*

Planar large-bipolarons can form if the ratio of the surrounding mediums' static to high-frequency dielectric constants is especially large, $\varepsilon_0/\varepsilon_\infty \gg 2$. A large-bipolaron in *p*-doped $La_2CuO_4$ is modeled as two electrons being removed from the out-of-plane orbitals of four oxygen ions circumscribed by four copper ions of a $CuO_2$ layer. These oxygen dianions relax inwardly as they donate electrons to the surrounding outwardly relaxing copper cations. This charge transfer generates the strong in-plane electron-lattice interaction needed to stabilize a large-bipolaron with respect to decomposing into polarons. The lowest-energy radial in-plane optic vibration of a large-bipolaron's four core oxygen ions with their associated electronic charges has *d*-symmetry. Electronic relaxation in response to multiple large-bipolarons' atomic vibrations lowers their frequencies to generate a phonon-mediated attraction among them which fosters their condensation into a liquid. This liquid features distinctive transport and optical properties. A large-bipolaron liquid's superconductivity can result when it undergoes a Bose condensation yielding macroscopic occupation of its ground-state. The synchronized vibrations of large-bipolarons' core-oxygen ions with their electronic charges generate this Bose condensate's dynamic global *d*-symmetry.

## I. Self-trapping and formation of a strong-coupling polaron

An electronic charge carrier becomes self-trapped in condensed matter when it is bound within the potential well produced by displacing the equilibrium positions of surrounding atoms from their carrier-free locations [1]. The corresponding strong-coupling polaron comprises the self-trapped electronic charge carrier together with the surrounding vibrating atoms. In addition, the stiffness constants governing these vibrations are generally reduced by the self-trapped carrier adjusting to the associated atomic displacements.

A strong-coupling polaron is described within the adiabatic approach [1]. Its wavefunction $\Psi(\boldsymbol{r}, \{\boldsymbol{R}\})$ is then the product of an electronic wavefunction as a function of atomic positions $\phi(\boldsymbol{r}, \{\boldsymbol{R}\})$ and the wavefunction describing atoms vibrating about displaced equilibrium positions $X(\{\boldsymbol{R}\})$:

$$\Psi(\boldsymbol{r}, \{\boldsymbol{R}\}) = \phi(\boldsymbol{r}, \{\boldsymbol{R}\})X(\{\boldsymbol{R}\}). \quad (1)$$

Here $\boldsymbol{r}$ designates the electronic carrier's position and $\{\boldsymbol{R}\}$ represents the set of atomic positions.

The adiabatic *limit* addresses polaron formation with the atoms presumed fixed at the displaced equilibrium positions they assume in the self-trapped electronic carrier's presence [1]. A polaron's net groundstate energy is the sum of its energy in this adiabatic limit $E_{ad}$ plus the softened vibrations' zero-point energy. A scaling analysis of polaron formation finds that $E_{ad}$ is the minimum of the energy of an energy functional $E_1(L)$ with respect to $L$, the parameter scaling the self-trapped carrier's spatial extent [1,2].

The energy functional associated with a self-trapped hole confined to a plane that is embedded within an ionic medium is [1,2]

$$E_1(L) = \frac{T}{2L^2} - \frac{U}{2L}\left(\frac{1}{\varepsilon_\infty} - \frac{1}{\varepsilon_0}\right) - \frac{V_p}{2L^2} = \frac{(T - V_p)}{2L^2} - \frac{U}{2L}\left(\frac{1}{\varepsilon_\infty} - \frac{1}{\varepsilon_0}\right). \quad (2)$$

Here $T$ and $U$ respectively denote electronic bandwidth and Coulomb interaction parameters. The second contribution describes the lowering of the potential energy resulting from displacements of the embedding medium's ions driven by their Coulomb interactions with the self-trapped electronic carrier. The ionic medium is characterized by its static and optical dielectric constants, $\varepsilon_0$ and $\varepsilon_\infty$, respectively. The third contribution, proportional to $V_p$, results from interactions of the self-trapped carrier with its plane's atomic displacements. As shown in Eq. (7) of Ref. (2), the contributions to this energy functional from intra-planar long-range and short-range electron-phonon interactions, proportional to $L^{d-4}$ and $L^{-d}$, respectively, both scale as $L^{-2}$ for $d = 2$. At its simplest this energy functional models a self-trapped hole confined to a $CuO_2$ layer in suitably doped $La_2CuO_4$.

When $V_p > T$ the self-trapped carrier collapses to the smallest physically meaningful size (e.g. a single site) thereby forming a *small* polaron [1]. Otherwise, the self-trapped carrier extends further and its polaron is referred to as being *large* [1].

A small polaron typically moves incoherently via a series of thermally-assisted hops. By contrast, a large polaron's motion is coherent since the change in its electronic energy as it moves between sites is generally less than the electronic transfer energy associated with this movement. Analogously, a large (singlet) bipolaron moves coherently whereas a small (singlet) bipolaron moves incoherently by thermally assisted hopping. Thus, only a large-bipolaron is viewed as a suitable basis for bipolaronic superconductivity [3-5].

## II. Formation of a large-bipolaron

A bipolaron forms when two electronic carriers of opposing spin occupy a common self-trapped state. The energy functional governing bipolaron formation differs from that for polaron formation in three respects. First, the sharing of a common potential well by a pair of carriers doubles their net kinetic-energy contribution: $T \rightarrow 2T$. Second, the bipolaron contribution to the self-trapping potential energy from two carriers occupying a common state in the self-trapping potential well is quadruple that for polaron formation since each of the two carriers shares a potential well whose depth is doubled: $2 \times 2 = 4$. Third, the two carriers experience their mutual Coulomb repulsion, $e^2/\varepsilon_\infty \left| \boldsymbol{r}_1 - \boldsymbol{r}_2 \right|$. Thus the energy functional for a planar bipolaron interacting with off-plane ions becomes [1,3-5]:

$$E_2(L) = \frac{2T}{2L^2} - 4\left[ \frac{U}{2L}\left( \frac{1}{\varepsilon_\infty} - \frac{1}{\varepsilon_0} \right) + \frac{V_p}{2L^2} \right] + \frac{U}{\varepsilon_\infty L} = 2E_1(L) + \frac{U}{\varepsilon_0 L} - \frac{V_p}{L^2}. \quad (3)$$

Equation (3) shows that off-plane electron-phonon interactions offset much of the mutual repulsion energy of a bipolaron's two carriers when $\varepsilon_0 >> \varepsilon_\infty$. Furthermore, as shown in Fig. 1, intra-plane electron-phonon interactions can foster the stabilization of a bipolaron with respect to two separated polarons. The minimum of $E_2(R)$ then lies below the minimum of $2E_1(R)$. However, the bipolaron will collapse into a small bipolaron if intra-plane electron-phonon interactions are too strong. Large-bipolaron formation is therefore limited by the condition [1,3-5]

$$\frac{4\,\varepsilon_0/\varepsilon_\infty - 6}{(\varepsilon_0/\varepsilon_\infty)^2 - 2} < \frac{2V_p}{T} < 1. \quad (4)$$

The window of acceptable values of $2V_p/T$ progressively opens as $\varepsilon_0/\varepsilon_\infty$ rises above 2. This condition appears fulfilled in the cuprate superconductors where $\varepsilon_0 \approx 50$ and $\varepsilon_\infty \approx 3$ [6-8].

Stabilization of a large bipolaron with respect to separating into two polarons also can be driven by reducing the paired electronic carriers' mutual Coulomb repulsion [1,9]. Electron correlation then lowers carriers' mutual Coulomb repulsion by keeping the two carriers away from one another. By contrast, the in-plane electron-phonon interaction drives large-bipolaron formation by enhancing the two electronic carriers' mutual overlap. A variational calculation incorporating both effects finds that a four-lobed large-bipolaron is at most only weakly stabilized by electron-correlation if the in-plane electron-phonon interaction is extremely weak; See Eq. (29) of Ref. (9). Otherwise, correlation effects are overwhelmed by those of the in-plane electron-phonon interactions. Hence, large-bipolaron formation is herein envisioned as being driven solely by the in-plane electron-phonon interactions.

### III. Condensation into a large-bipolaron liquid

There are additional physically significant polaron effects beyond the adiabatic *limit*. In particular, self-trapped electronic carriers' adjusting to the atomic vibrations lowers their stiffness constants [1,10,11]. For simplicity, consider Holstein's Molecular-Crystal Model, where each site consists of an harmonic oscillator with scalars describing its vibrational displacement and its stiffness [12]. Then, this polarization effect shifts oscillators' potential energy by:

$$V_{pol} = \frac{1}{2} \int d\boldsymbol{u} \int d\boldsymbol{u}' \left[ \sum_{\boldsymbol{g}} \Delta k_{\boldsymbol{g}}(\boldsymbol{u}, \boldsymbol{u}') \right] d(\boldsymbol{u}) d(\boldsymbol{u}'), \quad (5)$$

where $\Delta k_{\boldsymbol{g}}(\boldsymbol{u}, \boldsymbol{u'})$ denotes the reduction of the stiffness constant associated with oscillators located at $\boldsymbol{u}$ and $\boldsymbol{u'}$ generated by polarization of a singlet pair of self-trapped carriers centered at position $\boldsymbol{g}$. Here $d(\boldsymbol{u})$ and $d(\boldsymbol{u'})$ represent vibratory displacements of these oscillators from their carrier-induced equilibrium values. The reduction of a stiffness constant induced by the singlet pair of self-trapped carriers can be expressed as a superposition of matrix elements of the electron-phonon interaction between their orbital ground-state and their excited orbital states [10,11]. Denoting the $n$-th orbital eigenstate and corresponding electronic energy by $\left|\boldsymbol{g}, n\right\rangle$ and $E_n$ yields a simple expression for the carrier-induced reduction of the stiffness constant:

$$\Delta k_{\boldsymbol{g}}(\boldsymbol{u}, \boldsymbol{u'}) \equiv -2 \sum_{n \neq 0} \frac{\langle \boldsymbol{g}, 0 | Z(\boldsymbol{r}-\boldsymbol{u}) | \boldsymbol{g}, n \rangle \langle \boldsymbol{g}, n | Z(\boldsymbol{r}-\boldsymbol{u''}) | \boldsymbol{g}, 0 \rangle}{E_n - E_0}, \quad (6)$$

where $Z(\boldsymbol{r}-\boldsymbol{u})$ describes the magnitude and range of the electron-phonon interaction between a carrier at position $\boldsymbol{r}$ and an oscillator nominally located at $\boldsymbol{u}$. In particular, displacement of an oscillator at $\boldsymbol{u}$ by $\Delta(\boldsymbol{u})$ alters the electronic potential at $\boldsymbol{r}$ by $Z(\boldsymbol{r}-\boldsymbol{u})\,\Delta(\boldsymbol{u})$.

The ground-state energy of a collection of $n$ large bipolarons that are each centered at a site labeled as $\boldsymbol{g}$ is then given by [10,11]

$$E \cong -\sum_{g=1}^{n} \left( E_{\boldsymbol{g}, ad} + \varepsilon_{\boldsymbol{g}} \right) - \sum_{\substack{g'=1 \\ g' \neq g}}^{n} \sum_{g=1}^{n} \varepsilon_{\boldsymbol{g'}, \boldsymbol{g}}. \quad (7)$$

Here $-E_{\boldsymbol{g}, ad}$ denotes the ground-state adiabatic energy of a large bipolaron centered at site $\boldsymbol{g}$ produced by displacements of oscillators' equilibrium positions. In addition, adjustment of a site's self-trapped-carriers to these oscillators' vibrations lowers their zero-point energies by $-\varepsilon_{\boldsymbol{g}}$. The magnitude of this contribution to the shift of oscillators' zero-point vibration energies is primarily of first order in the fractional electronic carrier-induced shift of the stiffness constant, $k$:

$$\varepsilon_g \approx (\Delta k_g / k)\hbar\omega. \quad (8)$$

The collective contribution to the lowering of oscillators' zero-point vibration energy from self-trapped carriers centered at sites $\boldsymbol{g}$ and $\boldsymbol{g}'$ is represented by $-\varepsilon_{g'g}$. The magnitude of this coherent contribution is second order in the carrier-induced reduction of each site's stiffness constant:

$$\varepsilon_{g',g} \approx (\Delta k_{g'} / k)(\Delta k_g / k)F(|\boldsymbol{g}' - \boldsymbol{g}|)\hbar\omega. \quad (9)$$

The inter-(bi)polaron coherence factor $F(|\boldsymbol{g}' - \boldsymbol{g}|)$ falls as the ratio of the separation between large bipolarons to their radii $R$, $|\boldsymbol{g}' - \boldsymbol{g}|/R$, is increased, since self-trapped carriers interact primarily with phonons of wave-vector $\boldsymbol{q}$ that satisfy $|\boldsymbol{q}|R < 1$. By itself, this second-order contribution produces a phonon-mediated attraction between large bipolarons [1,5,10,11]. This effect is akin to the BCS phonon-mediated attraction between free electronic carriers [13,14].

The merger of bipolarons into grander polarons is precluded by the requirement imposed by the Pauli principle that some of its self-trapped carriers be promoted into excited states [1,5,10,11]. In other words, there is a short-range repulsion between bipolarons. In addition, since bipolarons move slowly in response to atomic movements, bipolarons repel one another at long range via their mutual Coulomb repulsions diminished by the material's static dielectric constant: $(2e)^2 / \varepsilon_0 |\boldsymbol{g}' - \boldsymbol{g}|$. Suppression of large-bipolarons' mutual long-range Coulomb repulsion with a sufficiently large static dielectric constant, enables the phonon-mediated attraction between large bipolarons to drive their condensation into a large-bipolaron liquid. Formation of such a liquid requires that the carrier density be modest enough so that large-bipolarons' formation is not precluded by their competing to displace the ions that surround them. This condition is plausibly fulfilled in hole-doped $La_2CuO_4$ where the firmly established maximum carrier density of the superconducting phase corresponds to having one singlet pair centered at about 10% of the unit cells of a $CuO_2$ layer [10,11]. Indeed, optical and transport experiments indicate that charge carriers in cuprates' normal state condense into a liquid [15-18]. This liquid of self-trapped singlet pairs is analogous to liquid $^4$He [1,5,10,11].

## IV. Planar large-bipolaron formation in cuprate superconductors' $CuO_2$ planes

As illustrated in Fig. 2, the core of a planar large bipolaron formed from two holes added to a $CuO_2$ plane of $La_2CuO_4$ is envisioned as involving four oxygen anions bounded by a square of copper cations. The hole-type bipolaron removes two of the eight out-of-plane electrons from these four oxygen anions ($2\times4 = 8$). In response the four oxygen anions will move radially inward toward one another while the surrounding copper cations relax outward.

Although oxygen's first electron affinity is negative ($-142$ kJ/mol), its second electron affinity is positive ($+780$ kJ/mol). Thus, the presence of nearby cations is required to bind an oxygen dianion's second electron. This outer-most electron is released from the oxygen dianion upon removal of the surrounding cations. More generally, electron transfers from dianions that reduce adjacent multi-valence cations increase as the dianion-cation separations increase. Thus, the outward relaxation of surrounding $Cu^{2+}$ cations away from the four-oxygen unit transfers electrons from oxygen anions toward $Cu^{2+}$ cations to convert them to $Cu^+$ cations. A singlet pair of holes forming this type of bipolaron is described by the chemical equation: 2 holes + $(O_4)^{8-}$ + $4Cu^{2+} \rightarrow (O_4)^{2-} + 4Cu^{1+}$. Then the core of this large bipolaron in hole-doped $La_2CuO_4$ contains a singlet pair of out-of-plane electrons distributed over four oxygen atoms.

This model of cuprates' $p$-type bipolarons produces distinctive Seebeck coefficients. The Seebeck coefficient is the entropy per carrier charge transported with an added carrier [1,19]. The large-bipolaron-induced reduction of the spin entropy arising from converting four spin-1/2 $Cu^{2+}$ ions into four spin-less $Cu^{1+}$ cations lowers the Seebeck coefficient of these large-bipolaron holes by an amount that increases with rising temperature. In the high-temperature paramagnetic limit the Seebeck coefficient of these $p$-type bipolarons garner a negative contribution of $-(k_B/2e)[4ln(2)]$, where $k_B$ denotes the Boltzmann constant and $e$ represents the elemental electronic charge's magnitude. This feature is consistent with the in-plane normal-state Seebeck coefficients of superconducting doped-$La_2CuO_4$ falling toward negative values after rising to a peak as the temperature is raised above the superconducting transition temperature [20-24].

## V. Vibrations of a cuprate large-bipolarons' core ions

The internal motions of this large-bipolaron's electrons are coupled to the radial vibrations of its core's four oxygen atoms. The adiabatic adjustment of the large-bipolaron's electrons to these atomic vibrations softens them and thereby tends to decouple them from other optic vibrations.

For simplicity, consider the radially symmetric vibrations of these four oxygen atoms of mass $M$ with the surrounding relatively heavy copper cations remaining fixed. The vibrational equations-of-motion governing radial displacements, the $r_i$, of these four oxygen atoms from their equilibrium positions are:

$$M\ddot{r}_1 = -Kr_1 - k_o(r_2 + r_4), \quad (10a)$$

$$M\ddot{r}_2 = -Kr_2 - k_o(r_3 + r_1), \quad (10b)$$

$$M\ddot{r}_3 = -Kr_3 - k_o(r_4 + r_2), \quad (10c)$$

and

$$M\ddot{r}_4 = -Kr_4 - k_o(r_1 + r_3). \quad (10d)$$

Here $K$ represents a vibrating oxygen ion's stiffness constant in the absence of displacements of neighboring oxygen anions. The stiffness constant $k_o$ depicts modulations of the restoring forces on an oxygen anion when nearest-neighbor oxygen anions are also radially displaced from their equilibrium positions. The modulations of oxygen anions' vibrations produced by other oxygen anions' vibrational displacements introduce dispersion of the frequencies of oxygen anions' optic vibrations. In fact, the above equations-of-motion are just the four-site analogue of the Molecular-Crystal Model which Holstein utilizes to describe the optic vibrations of an ionic solid [12].

The four mutually orthogonal normalized four-component eigenvectors of the four radial normal modes of these oxygen anions' vibrations are: $(1,1,1,1)/\sqrt{4}$, $(1,0,-1,0)/\sqrt{2}$, $(0,1,0,-1)/\sqrt{2}$ and $(1,-1,1,-1)/\sqrt{4}$. The four corresponding normal-mode vibrational frequencies are: $(\omega^2 + 2\omega_o^2)^{1/2}$, $\omega$, $\omega$ and $(\omega^2 - 2\omega_o^2)^{1/2}$, where $\omega^2 \equiv K/M$ and $\omega_o^2 \equiv k_o/M$. These normal-mode vibration frequencies correspond to progressively lower energies. As illustrated in Fig. 3, the highest-energy normal mode has $s$-like symmetry, the two degenerate normal modes have $p$-like symmetry and the lowest-energy normal mode has $d$-like symmetry. Furthermore, as shown in Fig. 3, this large bipolaron's two out-of-plane electrons alter their occupations of opposing oxygen atoms in response to their vibrations.

In analogy with Eq. (1), the collective ground-state associated with the Bose condensate of a large-bipolaron liquid comprises a nearly uniform distribution of large-bipolarons with their synchronized zero-point atomic vibrations. As represented in Fig. 4, the coherent in-phase zero-point vibrations of the central four atoms and their associated self-trapped electronic charge then displays a $d$-like symmetry. In other words, the phase of the Bose condensate possesses dynamic $d$-symmetry. This situation is exemplified in text-books where it is shown that the lowest-energy normal mode of coupled equivalent oscillators corresponds to their vibrating in-phase [25].

## VI. Discussion

Large-bipolarons' superconductivity describes the collective flow of its liquid in its coherently vibrating collective ground-state. Superconductivity is lost if large-bipolarons become pinned as they globally align commensurate with the underlying lattice rather than form a liquid. In hole-doped tetragonal $La_{2-x}Ba_xCuO_4$ two such compositions are close to the minimum and maximum concentrations for superconductivity. In particular, (1) the minimum hole concentration of the superconducting phase is close to 2/25 doping, one bipolaron per every (5×5) structural units and (2) the maximum doping is close to 2/9 hole doping, one bipolaron per every (3×3) structural units. Strikingly, superconductivity vanishes at the intermediate carrier density corresponding to $2/16 = 1/8$ doping, where there is one bipolaron per every (4×4) structural units [26].

The Bose condensation temperature of a large-bipolaron-liquid depends on the energy gap between its ground-state and its lowest-energy uniformity-destroying excitations. The lowest-energy collective excitations of a large-bipolaron liquid are plasma-like. The large-bipolaron-liquid's plasma frequency is $[4\pi n_{bp}(2e)^2/\varepsilon_0 m_{bpl}]^{1/2}$, where $n_{bp}$, $2e$ and $m_{bpl}$ respectively represent the density, charge and effective mass of the liquid's large bipolarons [27,28]. The small value of $n_{bp}$, the large value of $\varepsilon_0$ and the huge value of $m_{bpl}$ (since large-bipolaron motion requires significant atomic movement) relegate this frequency to a small fraction of the Debye frequency $\omega_D$, $\sim \omega_D(\varepsilon_\infty/\varepsilon_0)^{1/2}$. Furthermore the superconducting transition temperature varies with this energy $T_c \propto (n_{bp}/\varepsilon_0 m_{bpl})^{1/2}$. This feature explains several trends in superconducting doped oxides. Since $m_{bpl}$ is inversely proportional to the volume of a large bipolaron, $T_c$ increases as a $CuO_2$-based large bipolaron thickens to encompass progressively more contiguous $CuO_2$ layers. In addition, $T_c$ decreases with the addition of insulating layers between contiguous $CuO_2$ layers since $n_{bp}$ then decreases. Finally, the extremely large value of $\varepsilon_0$ of doped $SrTiO_3$ (20,000 at 1 K versus about 50 for cuprates) keeps its superconducting transition below 1 K [29-31].

The distinctive normal-state transport properties of a large-bipolaron liquid facilitate its identification. Large bipolarons and their liquid's excitations move with huge effective masses even more slowly than the sound velocity [27,28,32]. As a result, the scattering of large bipolarons and their excitations by long-wavelength acoustic phonons proceeds by their "reflection" as illustrated in Fig. 5. By contrast, the absorption or emission of acoustic phonons depicted in Fig. 5 dominates the scattering of conventional light-massed fast-moving charge carriers. For this fundamental reason, the rate with which large bipolarons and their excitations are scattered by long-wavelength acoustic phonons is less than even their Debye frequency $\omega_D$.

As illustrated in Fig. 6, large-bipolarons' Drude-like contribution to their frequency-dependent conductivity is therefore primarily limited to frequencies below $\omega_D$. In the dc limit the very long scattering time of a large bipolaron-liquid's excitations ($\tau > 1/\omega_D$) compensates for its huge effective mass to generate a moderate mobility, 1-10 cm$^2$/V-sec at 300 K, that remains inversely proportional to temperature even well below the Debye temperature [27,28,32-34]. Indeed, extremely large effective masses ($\sim$ 1100 free electron masses) are observed in cuprates but only at low enough frequencies to permit atomic motion (e.g. $< 10^{12}$ Hz) [6]. Furthermore, as measured in cuprates, the dc resistivity with a temperature-independent carrier concentration remains proportional to temperature even at very low temperatures [35-40].

Additional contributions to the frequency-dependent conductivity are produced above $\omega_D$ from optical absorptions that respectively liberate and excite large bipolarons' electronic carriers from and within their self-trapping potential wells. As shown in Fig. 6 this contribution consists of a broad asymmetric band plus some relatively narrow lower-frequency electronic absorptions [1,27]. The "gap" between the high-frequency electronic and low-frequency polaronic contributions to the frequency-dependent conductivity opens as the temperature is lowered thereby shifting the Drude-like contribution to lower frequencies [28,32].

Here the core of a large bipolaron in a *p*-type La$_2$CuO$_4$ superconductor is envisioned as two electrons removed from the out-of-plane orbitals on four planar oxygen dianions surrounded by four copper dications of a CuO$_2$ layer. Large-bipolaron formation proceeds with the inward relaxation of these four oxygen anions with the outward relaxation of the surrounding four planar copper cations. The lowest-energy optic-like vibrations of the four oxygen anions at the core of a large bipolaron and their associated electronic charge both possess *d*-symmetry. As a result, the synchronous in-phase vibrations of a liquid of these large-bipolarons imbue their Bose condensate with global dynamic *d*-symmetry.

Indeed, recent measurement of the temperature and doping dependences of the magnetic penetration depth and phase stiffness of a huge number of samples of overdoped La$_{2-x}$Sr$_x$CuO$_4$ imply that its superconductivity is attributed to charge carriers that form local pairs whose size is less than their characteristic separation [41]. Furthermore ultrafast spectroscopy measurements on some $La_{2-x}Sr_xCuO_4$ samples provide evidence of charge densities that fluctuate on picosecond time scales [42].

Photoemission and positron annihilation spectroscopies probe electron distributions by exciting and annihilating electrons encountered by incident photons and positrons, respectively. As enunciated by the Franck-Condon principal, these optical excitations and positron-annihilations occur on much shorter timescales than do atomic vibrations. Thus, each encounter takes a "snapshot" of a material's electronic and atomic distributions [43-45]. Above $T_c$, the electron-distributions associated with large bipolarons' self-trapped electronic carriers appear simply as a distribution of static localized centers. By contrast, measurements below $T_c$ find a uniform delocalized collective ground-state with its macroscopic $d$-symmetry.

Ion channeling is a relatively slow measurement that probes the passage of incident ions through a material. Above $T_c$, transiting ions are scattered by the incoherent vibrations of the host material's atoms. Below $T_c$, the coherent vibrations of the condensate's atoms facilitate the transit of ions through the host material [46,47].

The ratio of the static to optic dielectric constants $\varepsilon_0/\varepsilon_\infty$ is the crucial parameter in this theory. In particular, an exceptionally large value of this ratio ($>> 2$) is needed to form large bipolarons and their liquid. Large values of $\varepsilon_0/\varepsilon_\infty$ indicate an ionic solid with especially displaceable ions. Empirically, such materials are characterized by inter-ionic separations that differ from sums of the corresponding ionic radii. Fortunately, the ratio $\varepsilon_0/\varepsilon_\infty$ is directly measurable and inter-ionic separations are readily compared with tabulated common values.

Stabilization of large bipolarons with respect to decomposing into polarons is fostered by a strong-enough additional component of the electron-phonon interaction: c.f. $V_p$ of eqs. (3) and (4). Strong short-range electron-phonon interactions are associated with strong separation-dependencies of electron transfers from dianions to multi-valence cations. All told, ionic materials with unusually large values of $\varepsilon_0/\varepsilon_\infty$ ($>> 2$) that contain dianions and multi-valence cations are good candidates for large-bipolaron superconductivity. Cuprate superconductors are among these materials.

## Figure Captions

Fig. 1. Solid curves plot a bipolaron's energy functional $E_2(L)$ for three values of the in-plane electron-phonon interaction $V_p$ against the scaling length $L$. The dashed curve plots the energy functional for two separate polarons $2E_1(L)$ with $V_p = 0$ against the scaling length $L$. The adiabatic energies are at these curves' minima (indicated with arrows). Comparing the uppermost solid curve (for $V_p = 0$) with the dashed curve shows that the large bipolaron is unstable with respect to two separate large polarons. Increasing $V_p$ can stabilize large-bipolaron formation. The red curve shows the bipolaron collapsing into a small bipolaron if $V_p$ is too large, c.f. Eq. (4).

Fig. 2. The core of a $p$-type planar large-bipolaron on a $CuO_2$ layer in $La_2CuO_4$ is modeled. Without a bipolaron and its concomitant atomic relaxation this core comprises a square (shown with dashes) with $Cu^{2+}$ dications residing at its four corners and $O^{2-}$ dianions residing midway along its four edges. Removing two out-of-plane electrons from four oxygen dianions (green dots) shifts their equilibrium positions inward while surrounding copper dications (red dots) relax outward. Collaterally an electron is transferred from each of these oxygen *di*anions to the copper dications converting them into spin-less $Cu^+$ cations. The remaining two out-of-plane electrons (indicated in blue) occupy the oxygen anions.

Fig. 3. The symmetries and frequencies of quasi-radial normal-mode vibrations of the four oxygen ions (green solid circles) residing at the core of a $CuO_2$ layer's large-bipolaron are listed and schematically depicted. The concomitant changes of the electron-densities on these ions are indicated by the intensities of blue surrounding them.

Fig. 4. The lowest-energy synchronous vibrations of the core oxygen ions of large bipolarons in a $CuO_2$ plane imbue their Bose condensate with its overall dynamic $d$-symmetry. As represented in this figure, the large-bipolaron density is necessarily low.

Fig. 5. The strong scattering of a light fast-moving conventional electronic carrier produced by the absorption of a relatively large-momentum acoustic phonon (red arrow) depicted on the left is qualitative different from the very weak scattering of a heavy slow-moving large bipolaron generated by its "reflection" of a relatively small-momentum long-wavelength acoustic phonon depicted on the right.

Fig. 6. The frequency-dependent conductivity of a large-bipolaron liquid $\sigma(\Omega, T)$ has two distinct components. At applied frequencies $\Omega$ primarily below the Debye frequency $\omega_D$, large-bipolarons' collective motion generates temperature-dependent Drude-like contributions (depicted by blue curves). At frequencies well above $\omega_D$, optical liberation of large-bipolarons' self-trapped electronic carriers produces an asymmetric band (plotted in red). At somewhat lower frequencies, self-trapped carriers can be excited to states within their self-trapping potential wells (illustrated with several vertical red lines). The "gap" between the low- and high-frequency contributions opens as the temperature is lowered.

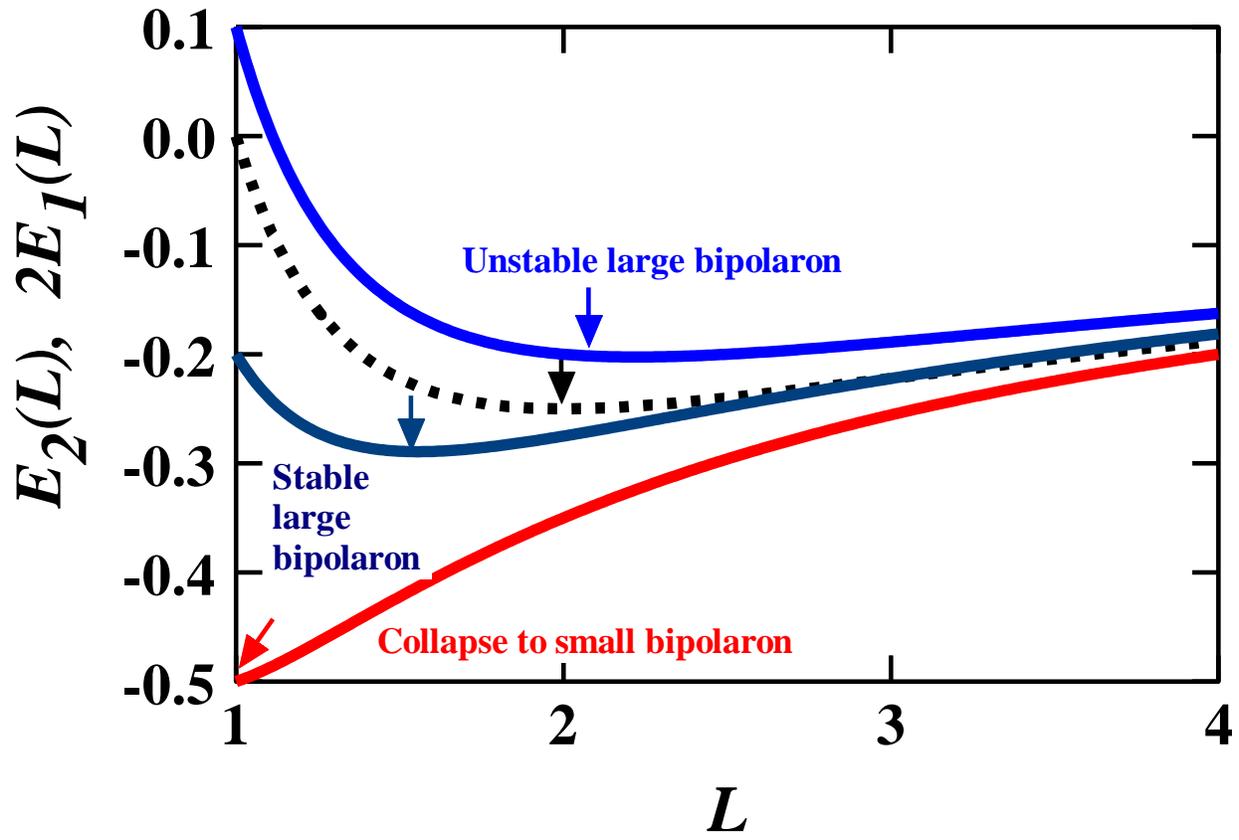

Figure 1

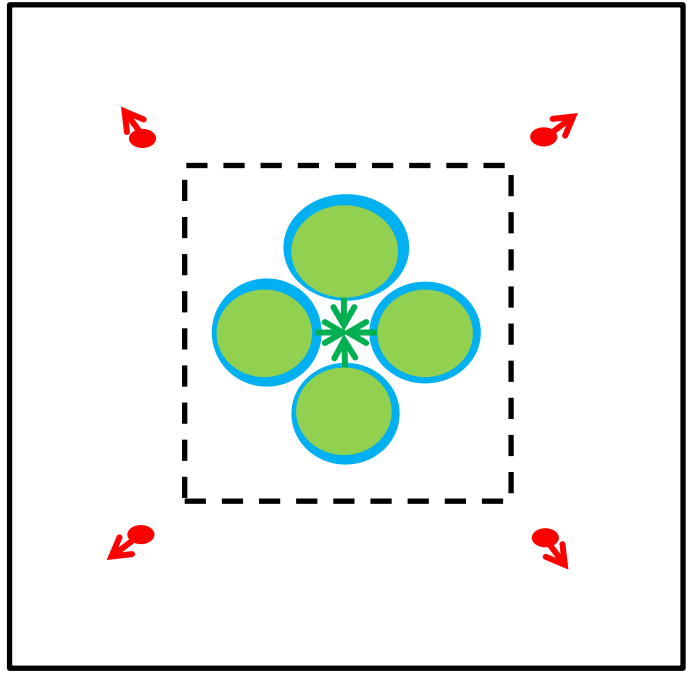

Figure 2

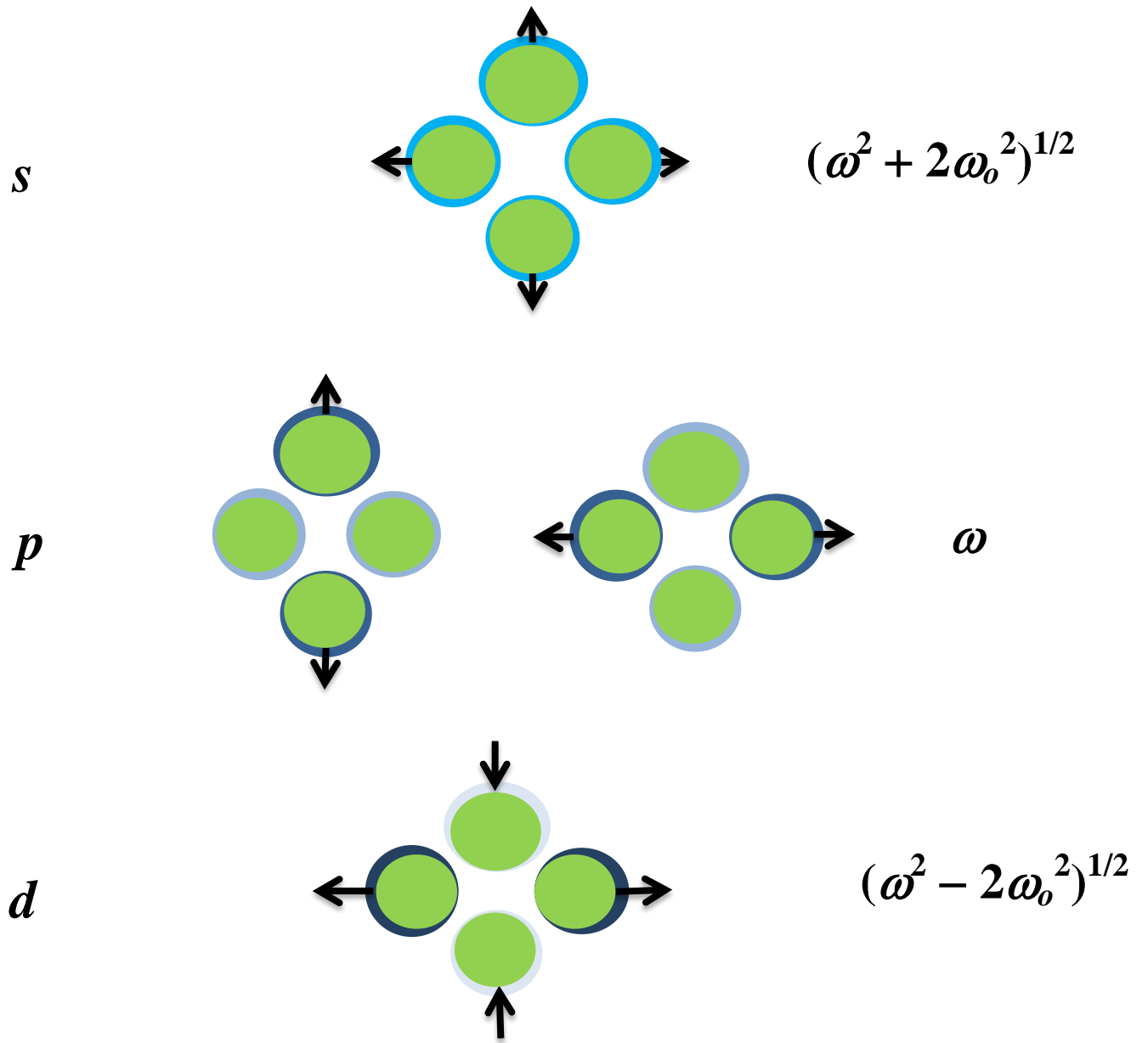

Figure 3

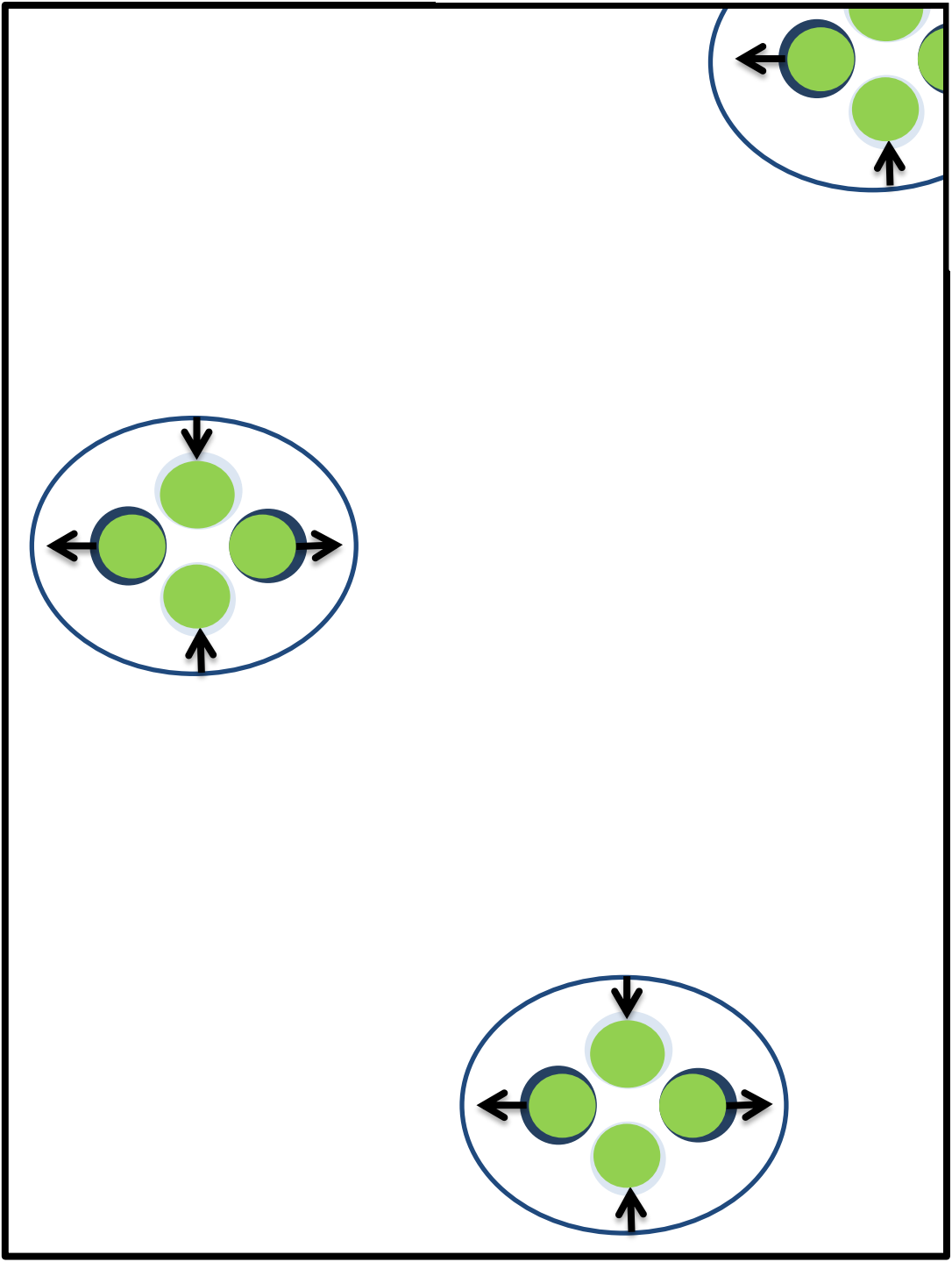

Figure 4

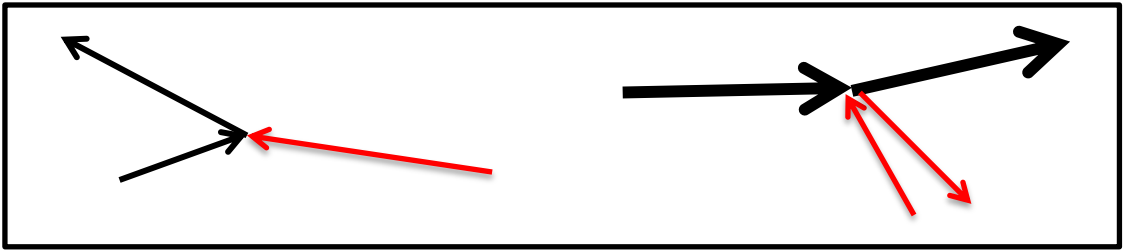

Figure 5

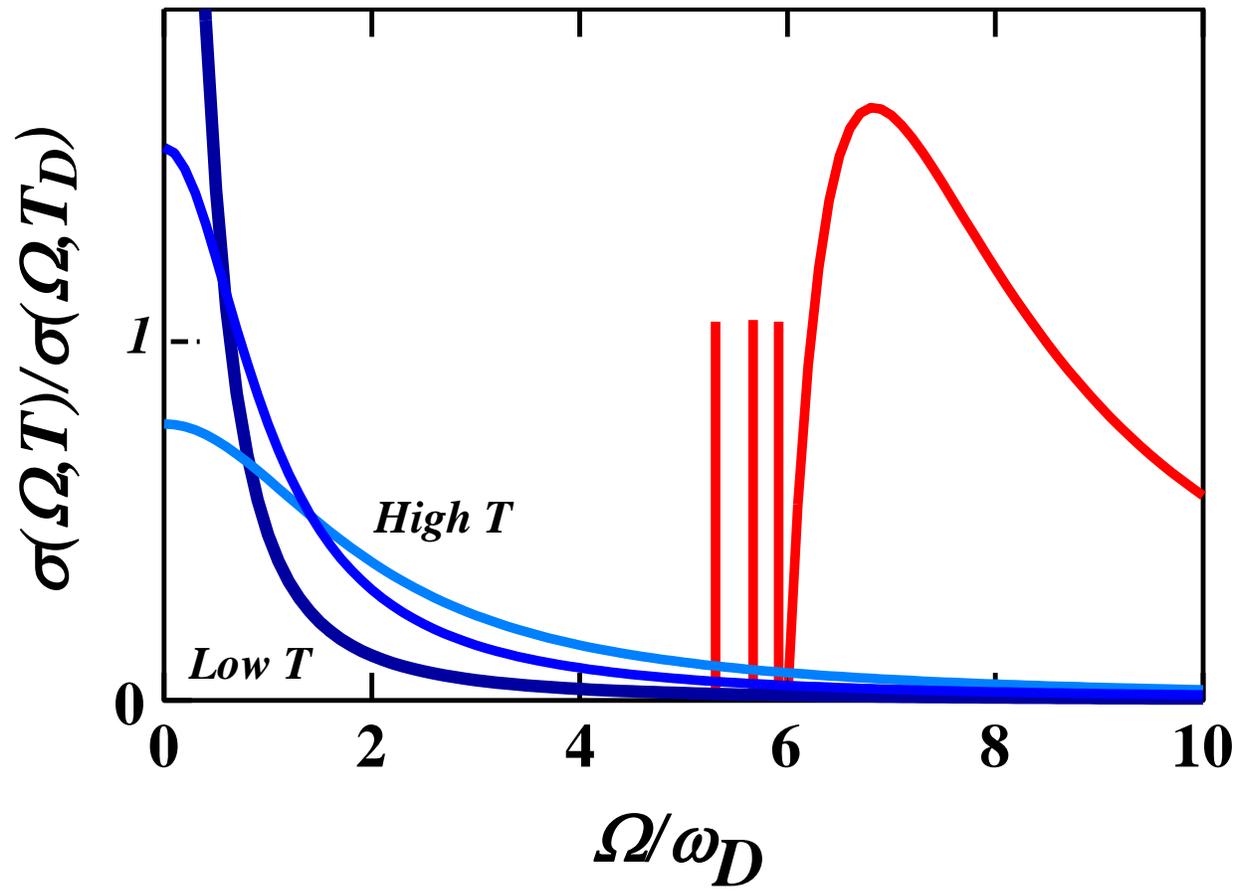